| | |
|---|---|
| Title: | Cathode to Target Simulations for Scorpius:I. Simulation Codes and Models |
| Author(s): | Ekdahl, Carl August Jr. |
| Intended for: | Report |
| Issued: | 2021-02-04 |

# Cathode to Target Simulations for Scorpius: I. Simulation Codes and Models

Carl Ekdahl

*Abstract*— The new Scorpius linear induction electron accelerator is under development for multi-pulse flash radiography of large, explosively-driven hydrodynamic experiments. Beam physics from the cathode to the target was examined with computer simulations, including envelope, ray-trace, and particle-in cell (PIC) codes. Beam instabilities investigated included beam breakup (BBU), image displacement, diocotron, parametric envelope, ion hose, and the resistive wall instability. Beam corkscrew motion and emittance growth from beam mismatch were also studied. The results of these simulations is documented in a series of reports. In this report the computer codes and physical models used for these simulations are described. The conclusion of this study is that Scorpius will produce and accelerate a beam with radiographic quality equivalent to the present accelerators at Los Alamos National Laboratory if the same engineering standards and construction details are upheld.

## I. INTRODUCTION

FLASH radiography of large explosively-driven hydrodynamic experiments is a time proven diagnostic in use world-wide [1, 2, 3] At Los Alamos National Laboratory two electron linear induction accelerators (LIAs) at the Dual-Axis Radiography for Hydrodynamic Test (DARHT) facility have provided bremsstrahlung radiation pulses for this purpose for more than a decade.

A new LIA called Scorpius is under development for flash radiography at the National Nuclear Security Site (NNSS) in Nevada [4]. Scorpius will use 102 solid-state pulsed-power (SSPP) driven 200-kV cells to accelerate a 1.4-kA electron beam from 2 MeV to 224 MeV in order to meet the following radiography requirements [5]:

- Four radiographic pulses within a 3-µs window
- Pulse length variable from less than 50 ns to 100 ns.
- Variable pulse spacing
- Capable of two pulses spaced within 200-ns center-to-center
- Radiographic resolution at least 0.8 lp/mm
.

The high-quality DARHT-I electron beam produces bremsstrahlung radiation source spots exceeding all anticipated requirements for hydrodynamic testing. Therefore, a new machine reproducing DARHT-I beam quality for multi-pulse radiography would satisfy mission requirements, However, there are significant differences between Scorpius and DARHT-I. For example, because of the lower injection energy and accelerating-gap voltages, and higher endpoint energy, there are 60% more cells in the Scorpius LIA that in DARHT-I, and the LIA is almost three times longer. Moreover, Scorpius will produce and accelerate four pulses, each of which is equivalent to the single DARHT-I pulse. These differences have a profound influence on beam stability.

Because of these differences, and the novelty of the Scorpius pulsed-power, an assessment of beam dynamic concerns is called for. These concerns include beam transport, motion, stability, and emittance, each of which was investigated in simulations that covered the entire birth-to-death lifetime of the beam; that is, from cathode to target (C2T).

For this assessment, we relied on analytic theory, simulation codes, and experimental data from the DARHT LIAs. The results of this assessment is the subject of a series of technical notes, each dealing with a particular aspect of beam quality. This, the first of these notes, presents an overview of the computer codes and physical models that we used, first for an early version of Scorpius based on traditional pulsed power [6], more recently for the present solid-state pulsed power (SSPP) design.

## II. SIMULATION CODES

Although a complete time-resolved C2T simulation can be performed with one of our envelope codes (LAMDA), we use several beam simulation codes to help us better understand beam dynamics in the DARHT LIAs. Most notably;

- Beam production in high-power diodes has been studied using the TRAK orbit-tracking and LSP particle-in-cell (PIC) codes.
- Beam transport and acceleration through the LIA has been studied using the XTR and LAMDA envelope and centroid codes, and the LSP PIC code,
- Coasting-beam transport from the LIA to the target has been studied using XTR, LAMDA, and LSP.

These are the four principal codes used to assess beam dynamics concerns, and to evaluate mitigating methods for Scorpius.

*A. TRAK*

TRAK is a finite-element ray-tracing code [7, 8, 9] using external fields generated by accompanying finite element electromagnetic codes [10, 11]. The program applies high-accuracy finite-element techniques to simulate steady-state beams in 2D (cylindrical) electric and magnetic fields, including effects of beam-generated electric and magnetic fields, and self-consistent space-charge-limited emission,

TRAK has been used to provide initial conditions for envelope code simulations of the DARHT beams, and has also been used as a diode design tool. For example, we have no beam measurements at the exit of DARHT-II diode, so we rely on predictions of TRAK and LSP to provide initial conditions for envelope codes. Fig. 1 illustrates how TRAK is used to provide the beam parameters at the diode exit, which serve as initial conditions for the beam simulations through the LIA. The location for hand-off to other codes was chosen to be far enough into the anode beam pipe that the applied diode electric field was reduced to less than 1% of the beam space-charge field, so it would have no influence on the transport simulations.

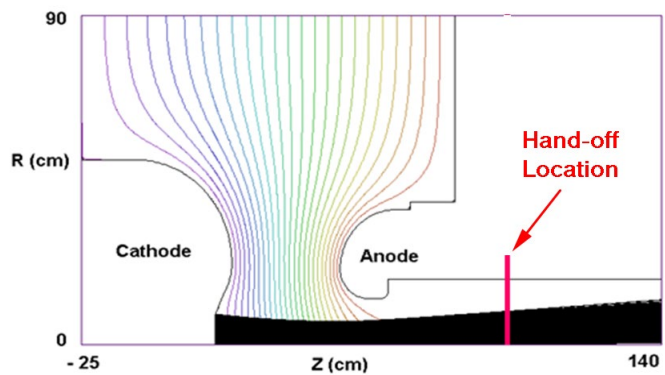

*Fig. 1: TRAK simulation of beam produced by the DARHT-II hot dispenser cathode diode. Equipotentials for the applied voltage are shown. The hand-off location shown is where beam current, energy, emittance, size, and convergence were extracted for use as initial conditions for further simulations with other codes. This hand-off location was chosen far enough into the anode pipe that the applied electric field was less than 1% of the beam space-charge field.*

An example of how TRAK has been used to guide design of new accelerators, such as Scorpius, is provided by the next illustration. Early on in the design of Scorpius there was a question of whether the beam produced by a 5-inch diameter cathode could be focused into a 6-iinch diameter beam pipe, which is significantly smaller than the 14-inch diameter DARHT-II pipe. A quick simulation by TRAK of a simple diode geometry with Pierce focusing electrodes laid this concern to rest in short order (Fig. 2).

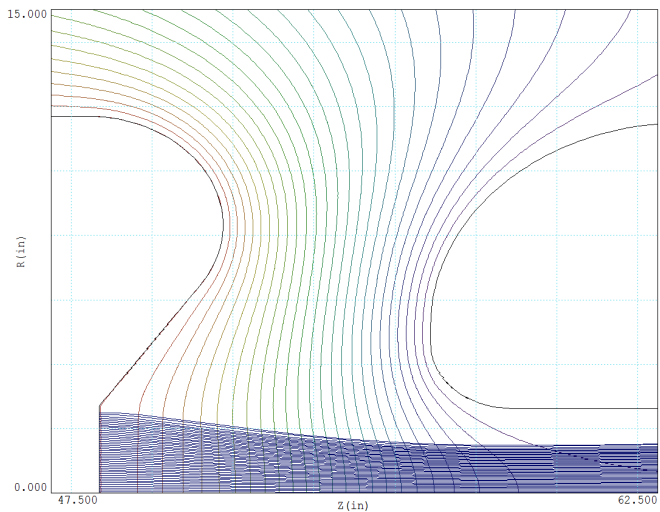

*Fig. 2: hypothetical 3-MeV, 2-kA beam produced by a 5-inch diameter cathode and focused into a 6-inch diameter beam pipe. Equipotentials shown in this figure include the beam space charge to illustrate the effect of the Pierce focusing geometry on the total field.*

More recently, TRAK has been used to design the Scorpius 2.0 MeV, 1.4-kA injector [12], and a 3-D version of the code (OmniTRAK) has been used to simulate the effects of cathode misalignment [13]. (These reports are also archived as Scorpius Technical Notes 027 and 036.) A recent TRAK simulation of the diode produced beam parameters at 1-m from the cathode that have been used for the cathode to target (C2T) simulations reported here (Fig. 3).

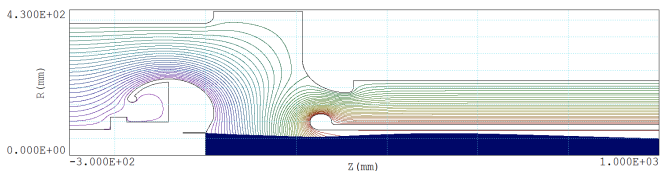

*Fig. 3: TRAK simulation of 1:1 push-pull diode producing a 2.0-MeV, 1.45-kA beam for hand-off to other codes at 100 cm from the cathode. At the hand-off position the beam envelope radius is 5.011 cm, with a 35.24 mrad convergence and a 204-mm-mrad normalized 4-rms emittance.*

*B. XTR*

XTR is a static envelope equation solver, and also a static solver of the beam centroid position. It is our primary tool for tuning the DARHT LIAs, as well as designing tunes for advanced LIAs such as Scorpius.

*1) Beam Envelope*

In XTR the radius of a uniform density beam is calculated from an envelope equation [14, 15, 16]. In the DARHT accelerators the beam is born at the cathode with no kinetic angular momentum. Moreover, a reverse polarity solenoid is used to cancel out the magnetic flux through the cathode so that the canonical angular momentum of the beam is zero. The envelope equation solved by XTR for such a beam is [17]

$$\frac{d^2 r}{dz^2} + \frac{1}{\beta^2 \gamma}\frac{d\gamma}{dz}\frac{dr}{dz} + \frac{1}{2\beta^2\gamma}\frac{d^2\gamma}{dz^2} r + k_\beta^2 r = \frac{K}{r} + \frac{\varepsilon^2}{r^3}. \quad (1)$$

It can be shown that this same equation holds true for any axisymmetric distribution undergoing self-similar current-distribution variations [14], where the radius of the equivalent uniform beam is related to the rms radius of the actual distribution by $r = \sqrt{2} R_{rms}$. Here, $\beta = v_e/c$, and $\gamma = 1/\sqrt{1-\beta^2}$, are the usual relativistic parameters. The beam electron kinetic energy is $KE = (\gamma - 1)m_e c^2$. Focusing by an axial magnetic field is characterized by the betatron wavenumber, which is

$$k_\beta = \frac{2\pi B_z}{\mu_0 I_A} \approx \frac{B_z(\text{kG})}{3.4\beta\gamma}\ \text{cm}^{-1}, \quad (2)$$

where $I_A = 17.08\beta\gamma$ kA, and $B_z$ is the axial magnetic field on axis. Corrections to magnetic focusing due to beam diamagnetism are calculated for a rigid rotor beam.

The two terms on the left involving derivatives of beam energy $\gamma$ are responsible for focusing the beam at the accelerating gaps. In XTR the gaps are approximated as thin Einzel lenses that increment the energy and focus the beam. In between gaps these terms give the envelope variation due to the beam space-charge potential depression. In XTR the potential depression is approximated by that for a uniform current density [15]

$$\varphi_{spd} \approx \frac{30 I_b}{\beta}\left[1 + 2\ln(b/r)\right] \quad (3)$$

where $b$ is the radius of the beam pipe.

Defocusing by the beam space charge is given by the first term on the right hand side of Eq. (1), and is characterized by the generalized perveance $K = 2 I_b / \beta^2 \gamma^2 I_A$.

Defocusing by beam transverse temperature is characterized by the beam emittance in the last term on the right hand side of the equation. The emittance that appears in Eq. (1) is

$$\varepsilon = 2\sqrt{\langle r^2 \rangle\left[\langle r'^2\rangle + \langle(v_\theta/\beta c)^2\rangle\right] - \langle rr'\rangle^2 - \langle rv_\theta/\beta c\rangle^2}, \quad (4)$$

which is related to the normalized emittance by $\varepsilon_n = \beta\gamma\varepsilon$.

In a solenoidal focusing system as in our LIAs the canonical angular momentum is conserved (Busch's theorem). The DARHT beams are born at the cathode with no mechanical angular momentum. Furthermore, the field angular momentum is zeroed there by using an opposite polarity solenoid to null the magnetic flux linking the cathode. Therefore, in the axial field of the solenoidal transport, the beam must rotate to conserve momentum, and for a uniform current distribution the beam rotates rigidly. For a rigidly rotating beam, the terms involving $v_\theta$ cancel, leaving simply

$$\varepsilon_n = 2\beta\gamma\sqrt{\langle r^2\rangle\langle r'^2\rangle - \langle rr'\rangle^2}. \quad (5)$$

In the azimuthal symmetry assumed in deriving Eq. (1), $\langle r^2 \rangle = \langle x^2 \rangle + \langle y^2 \rangle = 2\langle x^2 \rangle$ with similar expressions for the beam convergence and cross terms, so Eq. (5) reduces to

$$\varepsilon_n = 4\beta\gamma\sqrt{\langle x^2\rangle\langle x'^2\rangle - \langle xx'\rangle^2}, \quad (6)$$

which is the normalized Lapostolle "4-rms" emittance [18]. Differentiation of any of these equivalent expressions for the normalized 4-rms emittance shows that it is invariant in a system in which the forces on the beam envelope are constant or linear with radius at most. Even though solenoidal focusing fields and the space-charge field are generally nonlinear, the linear approximation is used for both.

The numerical method for solution is based on matrix optics [19] over incremental distances $dz$ short enough that $k_\beta$ can be assumed constant and the phase advance given by $k_\beta dz$. The interpolation scheme for advancing the solutions $r(z), r'(z)$ accounts for the diamagnetic and space-charge self-field corrections to the external solenoidal focusing, and the beam energy is incremented at the gaps.

Fig. 4 and Fig. 5 show the nominal tunes and beam envelope radii for the two DARHT accelerators as calculated with XTR.

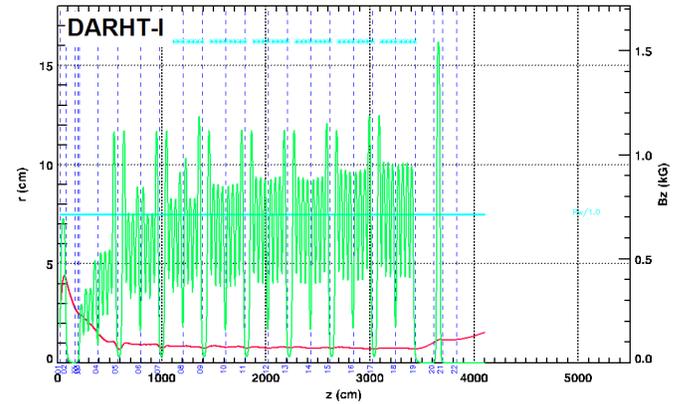

*Fig. 4: XTR simulation of beam transport through the nominal tune used on the short-pulse DARHT-I LIA. Green Curve: Magnetic field on axis produced by focusing solenoids. Red Curve: Beam envelope radius. Broken Blue Lines: Positions of BPMs. Cyan Asterisks 250-kV accelerating gaps. :*

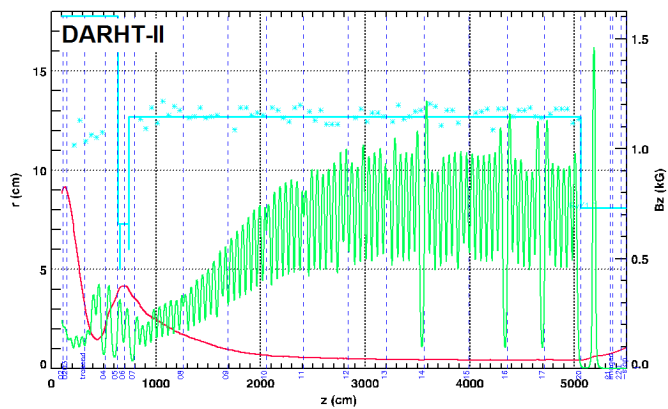

*Fig. 5: XTR simulation of beam transport through the nominal tune used on the long-pulse DARHT-II LIA. Green Curve: Magnetic field on axis produced by focusing solenoids. Red Curve: Beam envelope radius. Broken Blue Lines: Positions of BPMs. Cyan Asterisks 250-kV accelerating gaps. Cyan Line: Beam pipe wall and apertures:*

Even though the final energy of the DARHT-I beam is greater than the DARHT-II beam (19.8 MeV vs 16.5MeV), the LIA is shorter because the injected energy is greater (3.8 MeV vs 2.2 MeV), the gap voltage is greater (250-kV vs ~190 kV), less Volt-sec of magnetic material is needed to support the shorter pulses (<100 ns vs >1600 ns)..

XTR has been used to design tunes for beam transport through new accelerators. For example, Fig. 6 shows the XTR simulation results for a beam produced by the Scorpius injector and accelerated through the 12 cells planned for the integrated test stand (ITS). The beam was launched with parameters given in Fig. 3 at the hand-off position 100 cm from the cathode, transported through the anode beam-pipe using solenoidal fields, and then accelerated through the ITS cells with solenoidal transport fields low enough that the beam breakup instability gain could be measured.

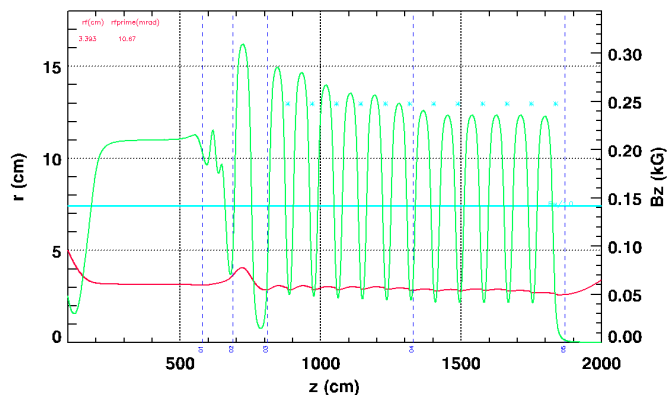

*Fig. 6: XTR simulation of beam transport through the injector anode pipe and the 12-cell ITS, where the 1.45-kA, 2.0-MeV injected beam is accelerated to 4.4 MeV.*

### 2) Beam Centroid Position

XTR can also calculate the position of the beam centroid as it responds to externally impressed solenoidal and dipole magnetic fields. The equations solved by XTR for the beam centroid positions $C_x(z), C_y(z)$ are

$$\frac{d^2 C_x}{dz^2} = k_\beta \frac{dC_y}{dz} + C_y \frac{dk_\beta}{dz} - k_y \qquad (7)$$

and

$$\frac{d^2 C_y}{dz^2} = -k_\beta \frac{dC_x}{dz} - C_x \frac{dk_\beta}{dz} + k_x \qquad (8)$$

where $k_x = B_x(kG)/1.7\beta\gamma$ cm etc., and includes dipole fields resulting from offset or tilted solenoids, as well as fields from steering dipoles. The second term in the right hand side of these equations is from the linear field approximation that retains only the first-order term of a Taylor series expansion of the solenoidal field that satisfies $\nabla \cdot \mathbf{B} = 0$; e.g., $B_x = -(x/2)dB_z/dz$ etc., where $B_z$ is the field on the axis. In XTR the beam energy $\gamma$ used in these centroid equations does not include space-charge depression, in order to agree with experiments [20].

An example of XTR beam centroid calculations is the simulation of beam motion through DARHT-II. Uncorrected beam motion at the exit of this LIA was dominated by an energy-dependent sweep. Since this would result in wandering of the radiographic source spots by more than their size, it had to be corrected, and XTR simulations provided guidance for the needed steering corrections. A major source of sweep was the folded current path in the injector. The diode beam source is at right angles to the coaxial current feed from the Marx generator [21, 22, 23, 24, 25, 26]. This asymmetry produces a weak transverse field in the diode anode-cathode gap that deflects the beam upward. Because of the defection the beam enters the solenoidal magnetic focusing field with an upward tilt, which causes the beam to follow a helical trajectory. This helix is initially large, and if uncorrected it remains large through the accelerator, as shown by simulations of the beam centroid position in Fig. 7. These simulations used measured beam centroid positions for initial conditions, magnetic field models fit to measurements for solenoidal and dipole fields, and measured solenoid misalignments for calculation of the resulting dipole fields. The predicted helical trajectory is stationary only if the beam initial energy and the cell accelerating potentials are constant in time. If either of these vary in time, the helix phase and gyro radius also vary at the LIA exit, causing the beam centroid position to sweep in time, as observed in experimental measurements.

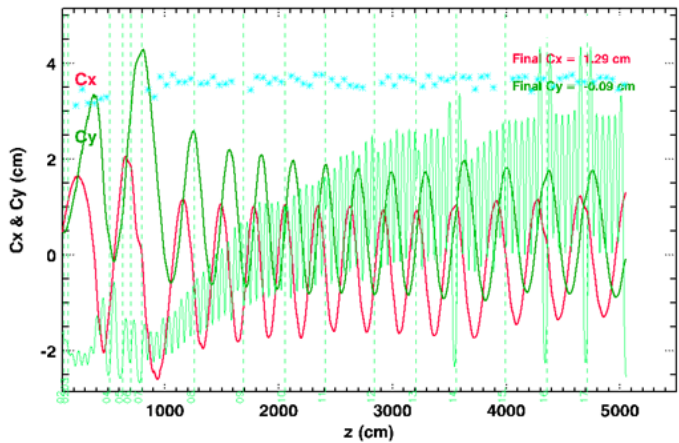

*Fig. 7: Simulation of the beam centroid position in x (Cx in red) and y (Cy in green). For this simulation the initial offset was uncorrected, and the resulting large helical trajectory extended through the accelerator. In this simulation the helical trajectory is stationary in time, because there is no time-variation of the initial energy or accelerating potentials.*

Obviously, the first step toward reducing beam motion at the accelerator exit would be to minimize the size of the helical trajectory caused by offset injection. Therefore, we corrected the offset injection by centering the beam using dipoles in the six injector cells. Fig. 8 is the result of an XTR simulation illustrating how this works by using two pairs of dipoles to correct the helical trajectory shown in Fig. 7. Experimental results were equally notable [24, 26].

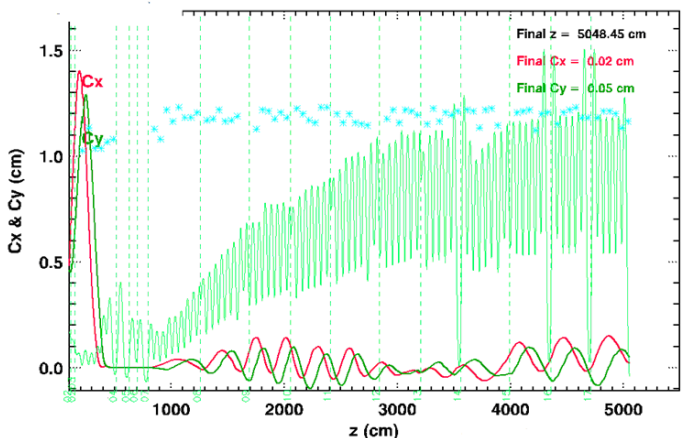

*Fig. 8: Simulation of the beam centroid position in x (Cx in red) and y (Cy in green). For this simulation the initial offset and resulting large helical motion has been corrected using 2 dipole pairs in the injector cells to center the beam (note trajectory between 300 cm and 800 cm). The residual helical motion in the main LIA is the result of dipole created by cell misalignments (measured misalignment values were used for this simulation).*

XTR is written in the IDL programming language, so it is easily customized by the user. Versions of the code exist that include features such as ad hoc emittance growth, exponential BBU growth, etc.

### C. LAMDA

LAMDA is an acronym for Linear Accelerator Model for DARHT. It is a time-resolved envelope and centroid solver, and it also has algorithms for calculating problematic beam instabilities. Moreover, LAMDA can simulate elliptical beams and quadrupole magnet focusing, which is important for tuning the downstream transport on DARHT-II.

LAMDA simulates time-resolved beam dynamics by subdividing the beam into many disks as shown in Fig. 9 [27, 28]. For a relativistic beam it is assumed that there is no interaction between the beam disks. The envelope radius for each disk is found from the envelope equation and the position of its center is calculated from the external forces.

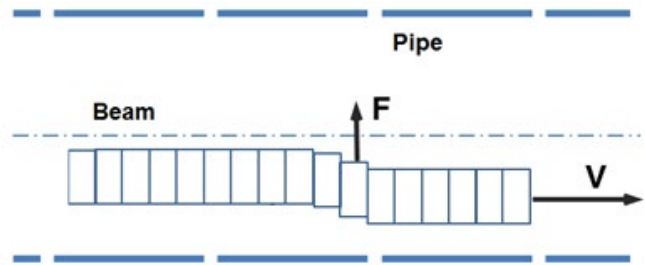

*Fig. 9: Articulated beam model used by LAMDA to simulate time resolved response of the beam envelope and centroid position to external forces.*

Focusing is calculated from files of axial magnetic field on axis and thin Einzel lens approximations for the accelerating gaps. For off axis fields, LAMDA uses the linear field approximation, which retains only the first order terms of a divergence-free Taylor expansion. Since normalized Lapostolle emittance is invariant for linear forces on the beam, it is held constant in LAMDA.

LAMDA includes models and algorithms for simulating many important beam instabilities:

- Beam breakup (BBU)
- Image displacement (IDI)
- Resistive wall
- Ion Hose

In addition, LAMDA includes calculations of solenoid misalignment dipoles enabling time-resolved simulations of the beam corkscrew motion that results from beam energy variations [29].

*1) Beam Envelope*

LAMDA uses an improved envelope equation that accounts for self-field terms [30] that are similar to those in XTR. These self-field terms include effects due to space-charge depression of the beam energy, diamagnetism from beam rotation, and the axial variation of the beam envelope. It is assumed that the current density is uniform, the beam rotates rigidly, and the divergence is radially linear.

A 4th order Runge-Kutta algorithm is used to integrate the equations for the beam envelope radius. Accelerating gaps are treated using either an advanced thin gap model [31], or by integrating through a user-supplied electric field map. In the thin-gap treatment, the beam energy is stepped by a constant over the pulse, varied in time per a user suppled file, or calculated from a circuit model.

*2) Beam Centroid Position*

To model the trajectory of the beam centroid, LAMDA solves the Lorentz force equation for each disk

$$\frac{dp_x}{dt} = e\left(E_x - \frac{v_z}{c}B_y + \frac{v_y}{c}B_z\right) ,$$
$$\frac{dp_y}{dt} = e\left(E_y + \frac{v_z}{c}B_x - \frac{v_x}{c}B_z\right) ,$$
(9)

and transforms the independent variable from time (*t*) to position (*z*) in the lab frame

$$\gamma\beta^2 x'' = -\gamma' x' + \frac{e}{mc^2}(E_x - \beta B_y + \beta y' B_z) ,$$
$$\gamma\beta^2 y'' = -\gamma' y' + \frac{e}{mc^2}(E_y + \beta B_x - \beta x' B_z) ,$$
(10)

where the prime symbol denotes $d/dz$, $\beta = v_z/c$, and $\gamma = \sqrt{1 - 1/\beta^2}$. The electromagnetic fields in these equations include all external fields (solenoids and gaps) plus the fields of the beam image in the beam pipe, including any fields responsible for instabilities. The full set of differential equations to be solved for the beam centroid motion are the equations for the spatial motion of each beam disk along the accelerator, and the equations for the temporal variation of the voltage at the gaps. A 4th order Runge-Kutta algorithm is also used to integrate the equations for the beam centroid.

Beam centroid motion simulations with LAMDA have been extensively used to investigate beam corkscrew motion [29]. They have been especially useful for long LIAs with phase advance so great that linear approximations for amplitude are invalid, because the growth is saturated. This saturation effect limits the growth of corkscrew in long LIAs, so the LAMDA studies have provided noteworthy guidance for the design of Scorpius [6], An example is illustrated in Fig. 10, which shows the beam centroid motion at the exit of an early version of Scorpius that used 72 gaps with each providing 250-kV of accelerating potential. This corkscrew motion resulted from a 2.4% rms variation of gap voltages interacting with the dipoles caused by random solenoid offsets, which were normally distributed with 0.29-mm rms misalignment.

As seen from Fig. 10, the rms amplitude is limited by the extent of the magnetic-flux enclosing cyclotron motion for a total phase advance exceeding $2\pi$, although for much less advance the amplitude could be linearly approximated by a short arc length. These LAMDA corkscrew simulations have helped quantify this saturation effect for long accelerators like DARHT and Scorpius. They have also helped evaluate the use of steering dipoles to significantly reduce the corkscrew amplitude via the "tuning-V" operational procedure [26, 32].

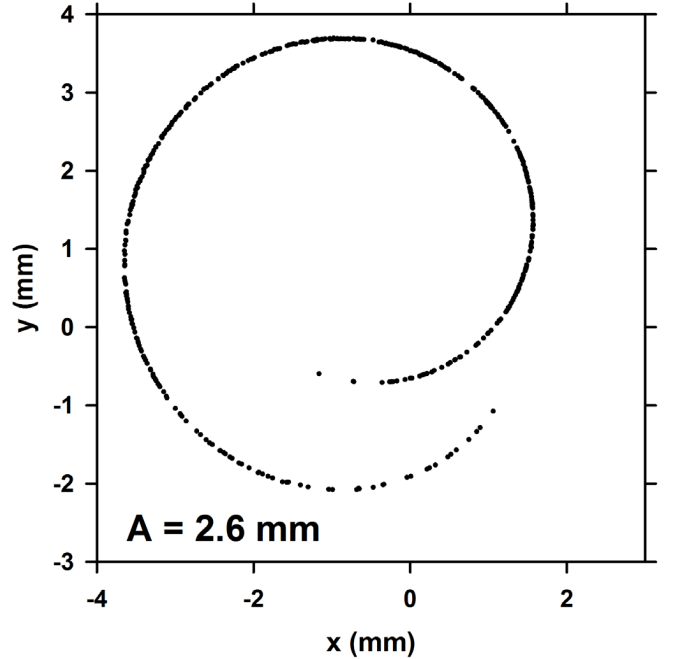

*Fig. 10: LAMDA simulation of beam-motion trajectory at the Scorpius LIA exit. Individual points are at 1-ns intervals, and represents the position of a single disk of the articulated beam model. This is the result of applying the voltage waveform with 2.4% rms fluctuations, and using the transverse fields calculated for 0.29-mm rms offsets of the solenoids.*

*3) BBU Algorithm*

The LAMDA BBU algorithm is based on Fourier transform methods introduced in seminal BBU publications [33].

The interaction of the disk with the dipole $TM_{1n0}$ fields at the gap is calculated using the narrow gap approximation. That is, the beam position is assumed be invariant as the beam crosses the narrow gap, but a transverse impulse is applied. LAMDA performs separate calculations of the transverse impulse in each of two orthogonal directions. A disk which arrives at a gap at time t (measured back from the head of the beam) receives a transverse impulse given by

$$\frac{\Delta p_x(t)}{mc} = \int_0^t \mathcal{Z}_x(t-\tau) \iota(\tau) \xi(\tau) c\, d\tau$$
$$\equiv \int_0^t \mathcal{Z}_x(t-\tau) f(\tau) c\, d\tau$$
(11)

and similarly for $\Delta p_y(t)$. Here, $\xi(t)$ is the disk displacement from the magnetic centerline in the x-direction, and $\iota(t) = I_b(t)/I_0$, where $I_0 = \frac{4\pi}{\mu_0}\frac{m_e c}{e} = 17.05$ kA. Also, $\mathcal{Z}_x(t)$ is the wake function for the cavity. Since the convolution integral is the Fourier transform of the product of the transforms of the integrand, this equation can be solved by using the inverse transform of that product. That is,

$$\frac{\Delta p_x(t)}{mc} = \int_{-\infty}^{\infty} Z(\omega) F(\omega) e^{-i\omega t} d\omega \qquad (12)$$

where $F(\omega)$ is the Fourier transform of $f(t)$, and the impedance $Z(\omega)$ is the transform of the wake function $\mathcal{Z}(t)$. Since the wake function is real, the real part of $Z(\omega)$ is symmetric, and $Z_\perp(\omega) = -iZ(\omega)$ to give a coupling impedance with antisymmetric real part that vanishes for zero frequency, as is usual for BBU theory [34].

Although the algorithm fully accounts for axial variation of the magnetic field as shown in Fig. 4 and Fig. 5, radial uniformity is assumed, and the radial increase of axial magnetic field in the focusing solenoids is ignored, as it is in the theory upon which the algorithm is based [33, 34]. This approximation has been shown to be justified in our LIAs [35]

Beam breakup simulated with LAMDA is illustrated by the following example. Consider an accelerator consisting of 100 cells, each with peak transverse impedance of 10 Ohm/cm at 800 MHz (Fig. 11). For this simulation the BBU was excited by a fast rising 2-kA pulse (Fig. 12) injected with a 1-mm offset. The rest of the parameters of the simulation are shown in Table I. In an LIA with a fast risetime current pulse such as shown in Fig. 12, BBU oscillations excited by the beam risetime grow to a peak and then decay [36, 34]. The LAMDA simulation of this RF motion at the exit of the LIA is shown in Fig. 13. The instability is convective; the peak moves back in the pulse as it propagates.

Table I. BBU Example Parameters

| Parameter | Symbol | Units | Value |
|---|---|---|---|
| **Beam:** | | | |
|   Kinetic Energy | $KE$ | MeV | 10 |
|   Current | $I_b$ | kA | 2 |
|   Pulse Risetime | $\tau_r$ | ns | 5 |
|   Pulse Flattop | $\tau_{flat}$ | ns | 60 |
|   Pulse Falltime | $\tau_f$ | ns | 4 |
|   Initial Offset | $x_0$ | mm | 1 |
| **Induction Cell:** | | | |
|   Resonant Frequency | $f_0$ | MHz | 800 |
|   Peak Impedance | $Z_\perp$ | $\Omega$/cm | 10 |
|   Quality Factor | $Q$ | | 4 |
| **Accelerator:** | | | |
|   Number of Cells | $N$ | | 100 |
|   Length | $L$ | cm | 8000 |
|   Pitch | $\ell$ | cm | 80 |
|   Uniform Guide Field | $B_z$ | kG | 0.4 |

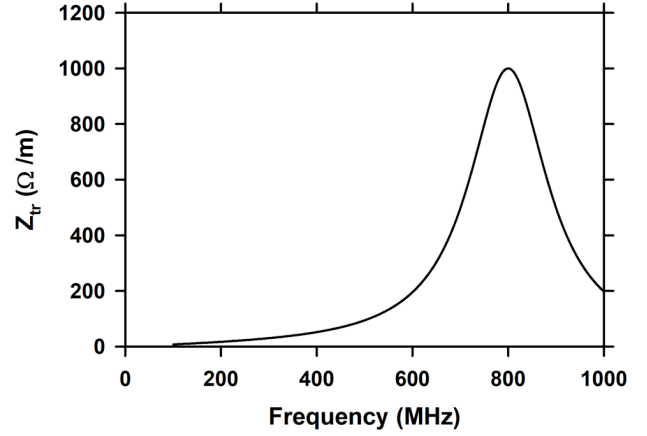

Fig. 11: Real part of transverse impedance that is responsible for the BBU instability. LAMDA wakefield model (Section ) using example parameters from Table I.

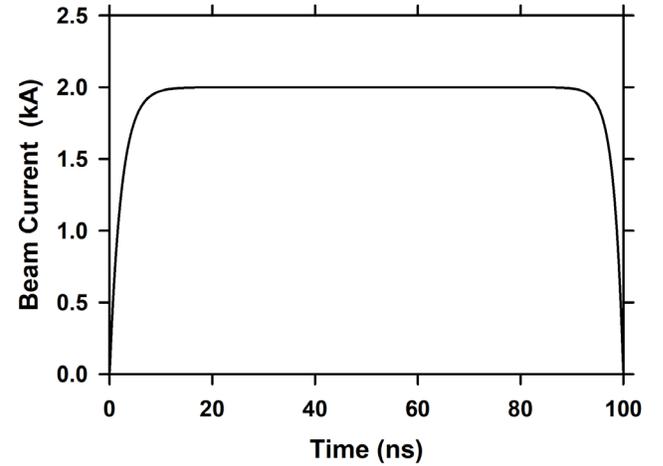

Fig. 12: Current pulse used for example simulation of BBU growth. For the example simulation this pulse was initiated with an offset of 1-mm in the x-direction.

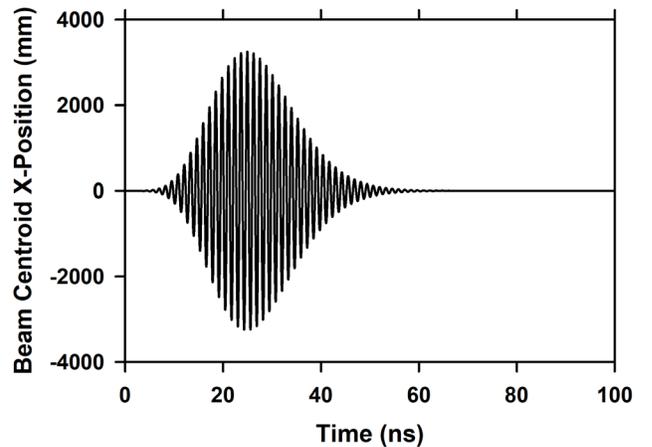

Fig. 13: Beam centroid motion resulting from BBU calculated by LAMDA for the example. The beam position in the horizontal plane at the LIA exit (z = 8000 cm) is shown.

Fig. 14 is a plot of the maximum amplitude of the displacement (over the beam pulse length) as a function of propagation distance. This displacement converges asymptotically to exponential growth predicted by theory [33, 37]. For a large enough number of accelerating cells the maximum BBU growth asymptotes to

$$\xi(z) = \xi_0 \left[\gamma_0 / \gamma(z)\right]^{1/2} \exp(\Gamma_m), \qquad (13)$$

where subscript zero denotes initial conditions, and $\gamma$ is the relativistic mass factor. The maximum growth exponent in this equation is

$$\Gamma_m(z) = \frac{I_{kA} N_g Z_{\perp \Omega/m}}{3 \times 10^4} \left\langle \frac{1}{B_{kG}} \right\rangle, \qquad (14)$$

where $I_{kA}$ is the beam current in kA, $N_g$ is the number of gaps, $Z_{\perp \Omega/m}$ is the transverse impedance in $\Omega/m$, $B_{kG}$ is the guide field in kG, and $\langle \ \rangle$ indicates averaging over $z$ [33, 34].

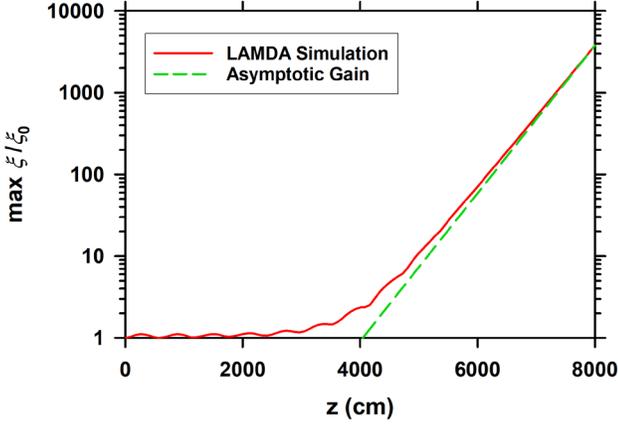

*Fig. 14: BBU growth in the example LIA. Red Curve: Growth simulated by LAMDA. Green Dashed Line: Asymptotic growth from Eq. (13) using parameters from Table I. The delayed start of strong BBU growth characteristic of some magnetic transport designs is clearly evident in this simulation.*

In actuality, the BBU amplitude grows enough that a significant portion of the pulse would be scraped off on the beam pipe long before the pulse transported to the exit. For example, according to Fig. 14, the beam oscillations would begin scraping on a 15-cm diameter pipe at about ¾ of the distance to the exit, and the pulse would be severely eroded before reaching the accelerator exit.

### D. LSP

The Large Scale Plasma (LSP) particle in cell code has been in use for several years at DARHT to simulate injectors and beam transport through the LIAs. DARHT injector simulations have been done in 2D (r, z) and in 3D (x, y, z) for representative diode-voltage pulsed waveforms, and these are ongoing for the Scorpius injector design. Similar spatial and time resolved simulations have been done to explore beam-target physics which contributes significantly to enlargement of the radiographic source spot, and to better understand some of the beam diagnostics used at DARHT or planned for Scorpius.

For beam propagation distances more than a few meters we use a slice algorithm [38] as a convenient means for avoiding the numerical Cerenkov instability, as well as speeding up the calculation. The instability is longitudinal, so limiting the distribution to a thin slice avoids it, and is an acceptable ansatz for a fully relativistic electron beam that has electromagnetic interactions collapsed to the transverse dimension. The LSP-Slice algorithm (LSP-S) has been used in 1D (radius) and 2D (x, y). It was used to better understand causes of emittance growth in DARHT-II [39, 26, 40, 41, 42], and it has been providing assessments of emittance growth for different architecture and tune designs for Scorpius [43, 44, 6, 45, 46].

The LSP-S algorithm is based on the LSP PIC code [38]. Initial electro- and magneto-static solutions are performed prior to the first particle push to establish the self-fields of the beam, including the diamagnetic field if the beam is rotating. After this initialization step, Maxwell's equations are solved on the transverse grid with $\partial/\partial z = 0$, and then the particles are pushed by the full Lorentz equations. At each time-step the grid is assumed to be located at the axial center-of-mass of the slice particles $z(t)$, which is propagating in the $z$ direction.

The initial particle distribution of the slice is either extracted from a full $x, y, z$ LSP simulation, or modeled as a uniform density rigid rotor with additional random transverse velocity (emittance). The rotation of the rigid-rotor model is consistent with zero canonical angular momentum in the given solenoidal magnetic field at the launch position, and the random transverse velocity is consistent with the specified emittance.

An example of agreement between LSP-S and XTR simulations of matched-beam transport and acceleration through DARHT-II is shown in Fig. 15. The slight differences that develop is attributed to non-self-similar evolution of the beam profile in LSP-S that cannot be accounted for in envelope calculations with XTR.

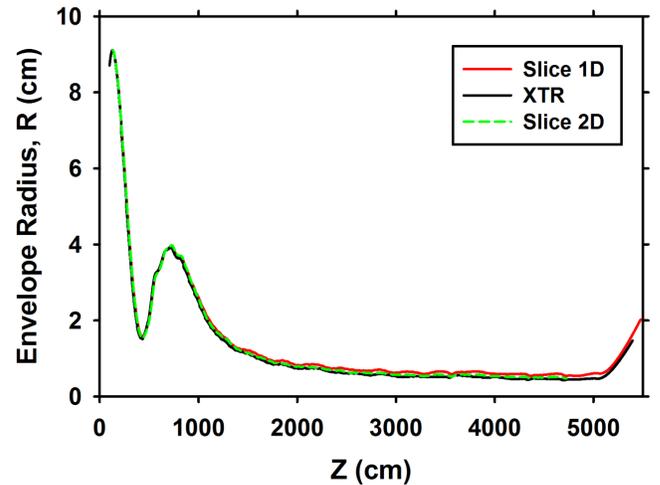

*Fig. 15: Comparison of the beam envelope through the DARHT-II LIA as calculated by the XTR envelope cade and the LSP slice algorithm (LSP-S) in cylindrical (1D) and Cartesian (2D) coordinates..*

Although the envelope equation only deals with axisymmetric beams centered on axis, the concept of beam emittance is much more general, and it can be calculated for non-axisymmetric distributions in LSP-slice simulations. Consider a beam with normalized distribution $\rho(x,x')$ in the $(x,x')$ plane of phase space. The position of the centroid of this distribution is at

$$\langle x \rangle = \iint x \rho(x,x) dx dx'$$
$$\langle x' \rangle = \iint x' \rho(x,x) dx dx' \qquad . \quad (15)$$

and the second moments are

$$\sigma_{xx} = 4\iint (x-\langle x \rangle)^2 \rho(x,x') dx dx' = 4\langle x^2 \rangle - 4\langle x \rangle^2$$
$$\sigma_{x'x'} = 4\iint (x'-\langle x' \rangle)^2 \rho(x,x') dx dx' = 4\langle x'^2 \rangle - 4\langle x' \rangle^2 \quad , \quad (16)$$
$$\sigma_{x'x} = 4\iint (x-\langle x \rangle)(x'-\langle x' \rangle)\rho(x,x') dx dx'$$
$$= 4\langle xx' \rangle - 4\langle x \rangle\langle x' \rangle$$

with $\sigma_{xx'} = \sigma_{x'x}$. One can define a 2x2 sigma matrix as

$$\boldsymbol{\sigma_x} = \begin{pmatrix} \sigma_{xx} & \sigma_{xx'} \\ \sigma_{x'x} & \sigma_{x'x'} \end{pmatrix} \qquad . \quad (17)$$

This matrix is related to the area occupied by the beam in the $x,x'$ projection of phase space by $A_x = \pi\sqrt{\det \boldsymbol{\sigma_x}}$ [47]. In beam optics theory the emittance is defined as $\varepsilon_{rms} = A/\pi$, so the rms emittance in the $x,x'$ cut through phase space is $\varepsilon_{x,rms} = \sqrt{\det \boldsymbol{\sigma_x}}$. Without loss of generality, one can center the beam in $x,x'$ space, and then has

$$\varepsilon_{x,rms} = 4\sqrt{\langle x^2 \rangle\langle x'^2 \rangle - \langle xx' \rangle^2} \quad , \quad (18)$$

which is the Lapostolle "4-rms" emittance [18]. Multiplying by $\beta\gamma$ gets the normalized emittance. The emittance calculated by LSP-S follows ref. [48] and generalizes to $\varepsilon_{rms} = (\det \boldsymbol{\sigma})^{1/4}$ [49], where $\boldsymbol{\sigma}$ is the $4 \times 4$ matrix formed from 4D moments as in Eq. (15) and Eq. (16) permuted through all transverse coordinates $x, x', y, y'$. This convention for $\varepsilon_{rms}$ reduces to Eq. (4) for axisymmetric beams.

The emittance calculated by LSP-S for the matched-beam transport shown in Fig. 15 is illustrated in Fig. 16. Here, it is seen that there is little, if any, emittance growth for a beam matched to the magnetic focusing field of a long LIA like DARHT-II. On the other hand, a beam injected with parameters different than those for which the magnetic tune was designed (mismatched) will undergo m=0 "sausage" oscillations with a long wavelength approximating the betatron wavelength in the moving average magnetic field and beam energy. This behavior is illustrated in Fig. 17 for a beam injected with an initial radius reduced by 20% from that required for matched transport. Betatron oscillations of the envelope are clearly evident, and are seen to be accompanied by a factor of three growth of emittance.

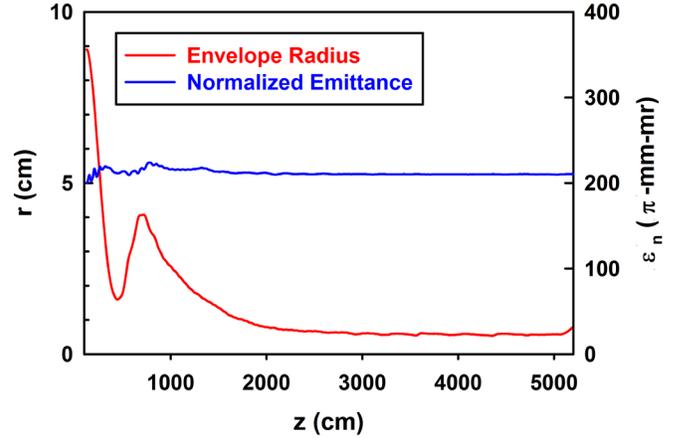

Fig. 16: Normalized beam emittance calculated by LSP-S PIC code for the matched-beam transport through DARHT-II shown in Fig. 15. (Red Curve) Beam envelope radius. (Blue Curve) Normalized 4-rms emittance.

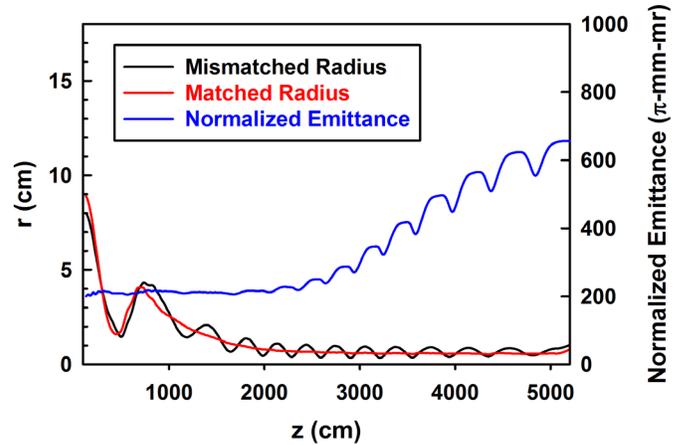

Fig. 17: Mismatched beam transport through the DARHT-II LIA. For this simulation the initial beam radius was decreased by 20% from the initial radius used for matched transport shown in Fig. 16. (Red Curve) Matched beam envelope radius from Fig. 16. (Black Curve) Mismatched beam envelop radius. (Blue Curve) Normalized 4-rms emittance of the mismatched beam.

Fully time-resolved LSP has been largely used to simulate beam production in the DARHT diodes, and also to model beam-target interactions that affect the radiography source spot. For example, Fig. 18 shows a sequence of snapshots of the beam produced by a hot dispenser cathode in the DARHT-II diode that illustrates the absence of beam spill in the oversized anode pipe during the long risetime of the pulse. As another example, Fig. 19 is a snapshot from a simulation of the beam focused onto a DARHT-I target showing acceleration of protons from beam-ionized gas desorbed from the target surface.

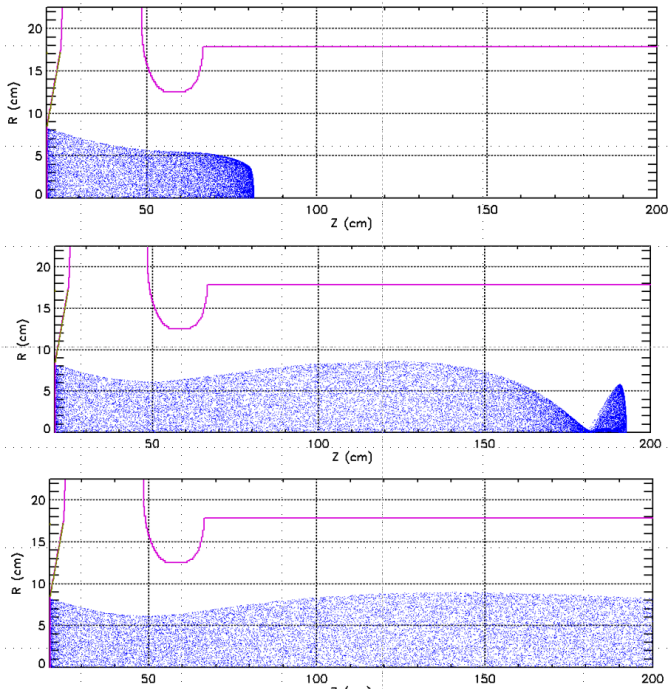

Fig. 18: Sequence of beam images from a DARHT-II diode simulation with LSP: (top) early in AK voltage risetime, (middle) late in risetime, (bottom) during flattop.

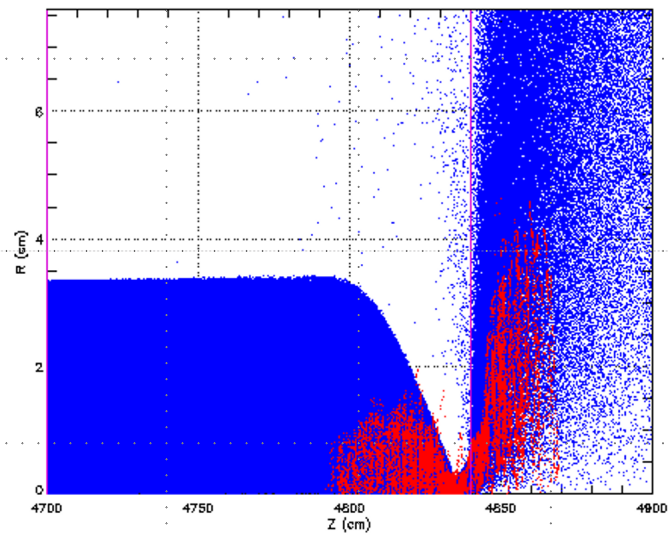

Fig. 19: Snapshot of beam interaction with target. Electrons are shown in blue, and protons are shown in red [50].

Plots of the transverse distribution of beam electrons obtained from 2D (Cartesian) LSP-S simulations have been particularly useful for revealing causes of emittance growth in the DARHT accelerators. For example, Fig. 20 shows the edge focusing due to cumulative spherical aberration in the DARHT-II injector and main LIA solenoids. Edge focusing has been theoretically established as a source of emittance growth [51], as well as direct spot size enlargement due to final focus aberration. Since spherical aberration is a strong function of beam size, we design the tunes for DARHT-II to rapidly focus the beam to a small size. Even though the edge focusing due to cumulative spherical aberration is noticeable in the simulations there is apparently little emittance growth from this effect in our simulations of matched beams (Fig. 16). However, cumulative edge focusing can account for the slightly larger radii at the DARHT-II exit calculated by LSP-S compared with XTR. The LSP-S distribution in Fig. 20 clearly has and rms radius that is larger than the rms radius of a uniform distribution with the same outer edge (XTR).

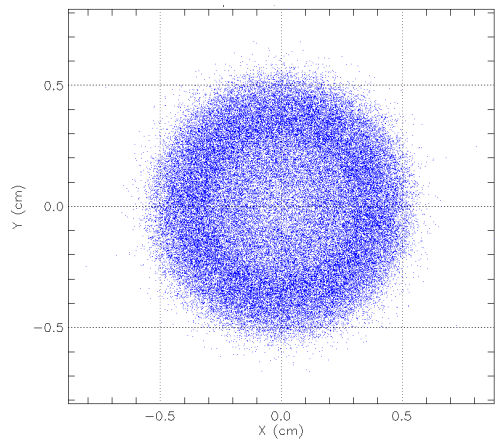

Fig. 20: LSP-Slice particle density in Cartesian coordinates showing the cumulative effect of edge focusing by solenoids from the diode to ~ 4200 cm (see Fig. 15). This plot has 0.5-cm grid-line spacing

Betatron oscillations of the envelope of a mismatched beam can parametrically pump halo growth through the particle-core mechanism [52]. This is the cause of the emittance growth shown in Fig. 17. This is revealed by comparing the development of halo on a mismatched beam with the growth of emittance, as illustrated in Fig. 21

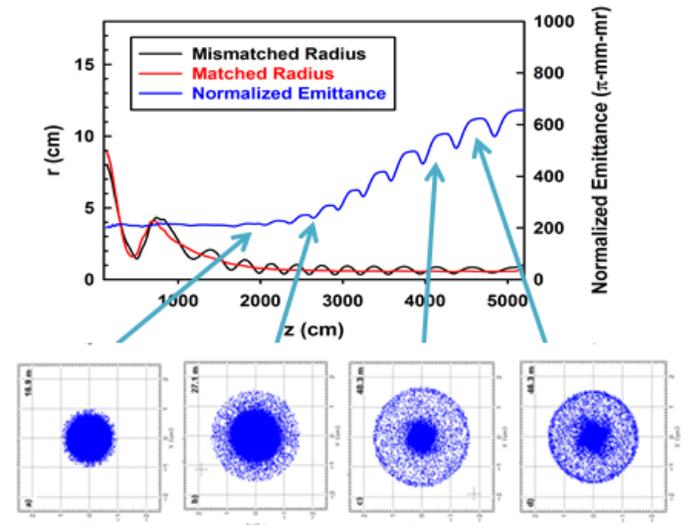

Fig. 21: Comparison of emittance growth with development of halo on a mismatched beam.

## III. Physical Models

### A. Magnetic Fields

In general, the external magnetic fields used by these beam dynamics codes are developed from either experimental data or output from field solvers such as POISSON or PerMag. Fields calculated by these solvers have been extensively compared with physical measurements and found to be in agreement.

Examples of magnetic field simulations used for TRAK, LSP and LAMDA are shown in the next few figures. Fig. 22 shows the magnetic field in the DARHT-I diode as calculated by PerMag [10, 11] for TRAK [7], and Fig. 23 show the field in the DARHT-II diode as calculated by PerMag. The field of a cell solenoid for an advanced LIA design (ARIA [43]) is shown in Fig. 24.

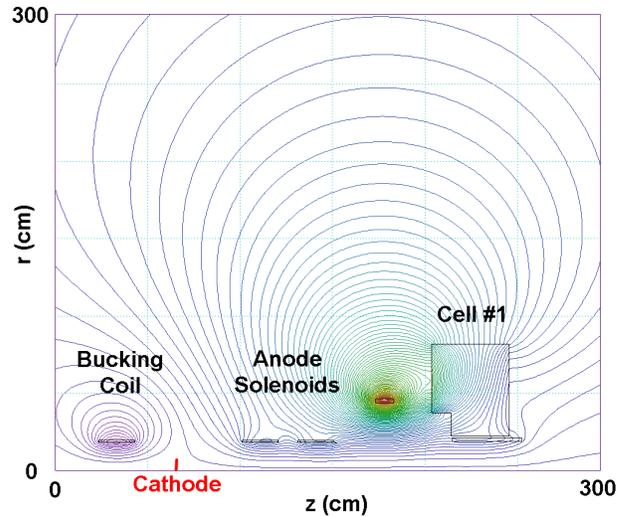

Fig. 23: Magnetic field in DARHT-II diode region as calculated by PerMag for TRAK simulations of the injector. For DARHT-II, the injector simulations include one or more injector cells, which included the Metglas cores as shown here.

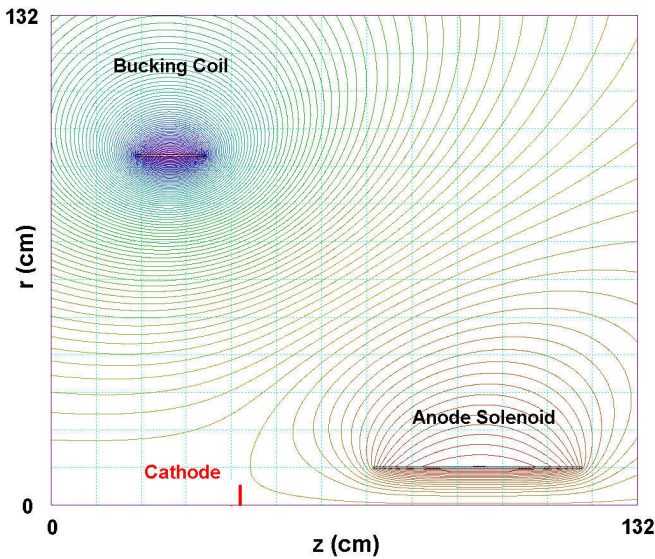

Fig. 22: Magnetic field in DARHT-I diode region as calculated by PerMag for TRAK simulations of the injector. The location and extent of the explosive-emission velvet cathode is shown in red. The purpose of the reversed field bucking coil is to null the canonical angular momentum by cancelling the flux linking the cathode. Tabular field maps from this or similar simulations can be used as input to LSP. Alternatively, LSP can use a simple table of the field on axis.

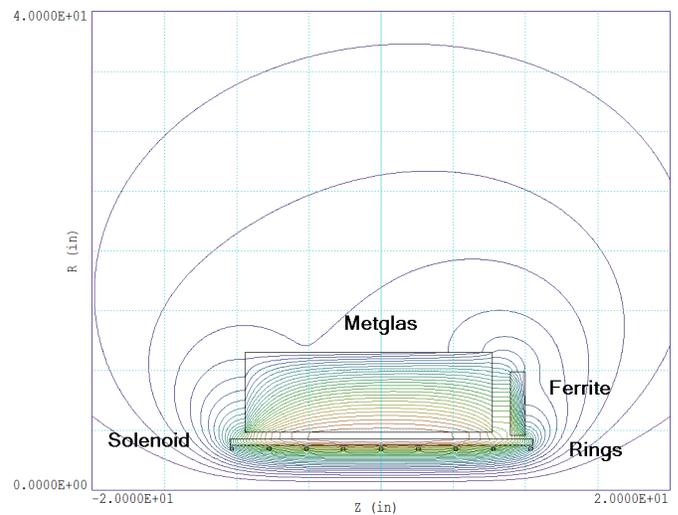

Fig. 24: Magnetic field of cell solenoid calculated by PerMag for an early design of an advanced radiography accelerator driven by conventional pulsed power (ARIA [43]). This simulation included both Metglas cores and a ferrite disk used to replicate the RF boundary conditions of the DARHT-I cavity. Also shown are the steel homogenizing rings used to suppress unwanted multipole fields resulting from winding asymmetries. The on-axis field of individual solenoids such as this are fit by the model given by Eq. (19) for use with XTR or LAMDA, which can also use files created directly from such simulations in order to include gross asymmetries. The fields of individual solenoids are superimposed to create the field of an LIA for input to LSP.

The envelope codes require the magnetic field on axis. For XTR, solenoidal focusing fields are calculated from a model generalized from the field of an ideal single-turn current sheet. The XTR solenoid field on axis is modelled by [17]

$$B(z) = B(0)\exp(-\alpha z^2) \times \left[\frac{\left[(L/2)^p + R^p\right]^{1/p}}{L}\right] \times \left\{\frac{L/2-z}{\left[(L/2-z)^p + R^p\right]^{1/p}} + \frac{L/2+z}{\left[(L/2+z)^p + R^p\right]^{1/p}}\right\} \quad (19)$$

Here, $L$ and $R$ are the effective length and radius of the solenoid, not necessarily the physical dimensions, and the parameter $p$ is exactly 2 for an ideal sheet current. Also, $B(0)$ is the maximum field given by $B(0) = C_B I_{mag}$, where $I_{mag}$ is current powering the solenoid, and $C_B$ is in G/A. The parameter $\alpha$ is used to sharpen the profile in order to better model pole pieces, if needed.

The five parameters required as input to XTR are determined by fitting to measurements or simulations. The solenoid locations are specified in an XTR input file of the lattice, and the LIA transport is built up by applying Eq. (19) at each location. Dipole steering fields for XTR centroid calculations follow the same formulation for axial variation as the solenoidal fields

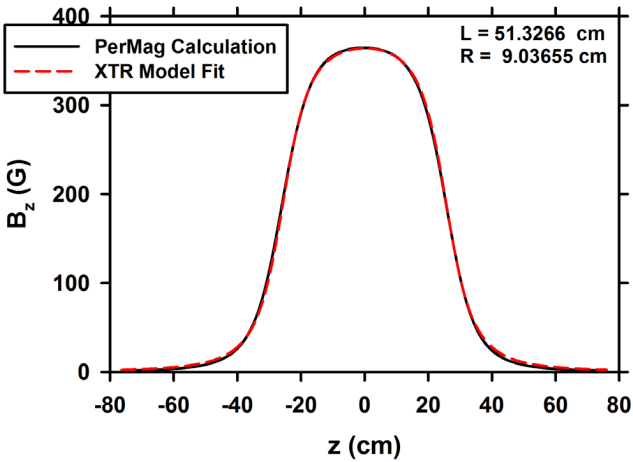

Fig. 25: Comparison of the magnetic field on axis calculated by PerMag for the ARIA cell design shown in Fig. 24 and the XTR model (Eq (19). ) fitted to it by maximum likelihood. For asymmetry as slight as shown here, the field calculated by the XTR model is also used by LAMDA

.
Since LAMDA has the capability for simulating elliptical beams, it requires all three field components on axis, which are input as files that also include transverse first derivatives needed for the equation of motion solver.

For PIC simulations with LSP, external fields are input as functions of $z$ using either full 3D maps or tables of values on axis, and are applied at the instantaneous axial center-of-mass location. External fields that are azimuthally symmetric, like those from solenoids and gaps, are input as tables of on-axis values. The off-axis components are calculated up to sixth order using a Taylor series expansion that satisfies the Maxwell equations [16];

$$B_z(r,z) = B(z) - \frac{r^2}{4}B''(z) + \frac{r^4}{64}B''''(z) - \cdots$$
$$B_r(r,z) = -\frac{r}{2}B'(z) + \frac{r^3}{16}B'''(z) - \cdots \quad (20)$$

where the magnetic field $B(z)$ is the field on axis, and the prime symbol designates differentiation with respect to the axial coordinate, $z$. Using this expansion, spherical aberrations of the accelerator optics are included in the slice simulations to the order of expansion. The result of accumulating high-order spherical aberration through the DAHRT-II LIA is illustrated in Fig. 20.

Transverse magnetic fields from steering dipoles and cell misalignments are input as $x, y$ values that uniformly fill the solution space, an approximation that is obviously best for a beam close to the axis. These dipole fields were obtained from XTR, which calculates them on axis from steering dipole excitation currents and cell misalignments, which have been measured [53, 54].

B. Electric Fields

For electric accelerating gap fields XTR simply increments the beam energy, and focuses with a thin lens model for the Einzel lens of the gap [17]. LAMDA uses a more sophisticated thin lens model [31], and also admits files of the field on axis [28]. The voltage applied to a gap in LAMDA can be time varying, and either read from a file, or calculated from available built-in pulse models. Electric fields for LSP are provided as full maps, or just the field on axis, from which off axis components are derived with a Taylor expansion like Eq. (20).

The files needed for external electric fields in these codes are created from the output of field-solver codes like POISSON or Estat [10, 11]. Fig. 26 shows the equipotentials of the DAHRT-I gap electric field calculated by Estat. This semi-shielded gap design is also used on Scorpius.

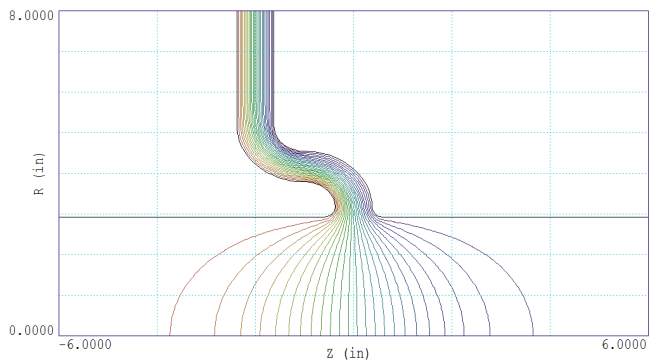

*Fig. 26: Electric accelerating potentials in the beam pipe as calculated by Estat for the DARHT-I cell gaps.*

### C. Transverse Impedance

As discussed earlier, the LAMDA BBU algorithm requires the transverse impedance of the induction cells as a function of frequency to compute BBU growth. These have been calculated using AMOS [55, 56] or CST [57], and/or measured [57, 35]. The data from these calculations and measurements has been used to determine the parameters of the LAMDA impedance model.

The wakefield transverse impedance used in LAMDA is [58, 59, 60]

$$Z_\perp = \sum_n Z_{\perp n} \left[ \frac{1}{1 + 2iQ_n(\omega/\omega_n - 1)} - \frac{1}{1 + 2iQ_n(\omega/\omega_n + 1)} \right] \quad (0.21)$$

Because theoretical expressions for transverse impedance vary in the phase relation between real and complex components depending on author, it is useful to remember that the component responsible for BBU coupling is real, satisfies $Z_\perp(-\omega) = -Z_\perp(\omega)$ (antisymmetric), and vanishes at the origin [34]. The real part of a single frequency impedance using this model with parameters in Table I is shown in Fig. 11. LAMDA has the capability for using different impedances for vertical and horizontal polarizations to account for the known asymmetries of the DARHT cells. Moreover, the code has recently been revised to accommodate up to twelve different resonant frequencies, in order to better represent the fine structure of the impedance of the Scorpius cells.

## IV. CONCLUSIONS

Beam physics for the new Scorpius radiography accelerator has been informed by simulations with several codes, four of which have been briefly described here. Although it is entirely feasible to do cathode to target (C2T) simulations with a single code, such as the LAMDA time-resolved envelope code, we have found it advantageous use several codes in order to provide a more accurate assessment of beam dynamics issues. In short, we use the code most suited to solving the problem in question.

Using this strategy for Scorpius, we have found that, if the engineering standards used on the DARHT accelerators are adhered to, there should be little difficulty with beam dynamics. Based on these simulations and calculations, we do not expect disruptive instabilities or excessive emittance growth in that novel accelerator, or indeed in *any* radiography accelerator based on DARHT-I.

Although this report has concentrated on explaining the major computational tools used to better understand DARHT beam physics and design new radiography LIAs, it should be noted that they have been often supplemented by other beam dynamics codes, including some written in house.

### ACKNOWLEDGMENTS

The author is indebted to his colleagues in the J-6 DARHT Group at Los Alamos National Laboratory, and in the multi laboratory ASD-Scorpius Project, for stimulating discussions on these and related topics.

This work has been supported by the National Nuclear Security Agency of the US Department of Energy under contract number 89233218CNA000001.

### REFERENCES

[1] T. H. Martin, A. H. Guenther and M. Kristiansen Editors, J. C. Martin on pulsed power, Springer, 1996.

[2] C. Ekdahl, "Modern electron accelerators for radiography," *IEEE Trans. Plasma Sci.*, vol. 30, no. 1, pp. 254-261, 2002.

[3] K. Peach and C. Ekdahl, "Particle radiography," *Rev. Acc. Sci. Tech.*, vol. 6, pp. 117 - 142, 2013.

[4] M. T. Crawford and J. Barraza, "Scorpius: The development of a new multi-pulse radiographic system," in *Proc. 21st IEEE Int. Conf. Pulsed Power*, Brighton, UK, 2017.

[5] M. Crawforrd, "ASD-Scorpius overview," Los Alamos National Laboratory Technical Report LA-UR-20-26293, 2020.

[6] C. Ekdahl, "Beam dynamics for the Scorpius Conceptual Design Report," Los Alamos National Laboratory Technical Report, LA-UR-17-29176 and ArXiv:1710.11610, 2017.

[7] S. Humphries, "TRAK: Charged particle tracking in electric and magnetic fields," in *Computational Accelerator Physics*, R. Ryne, Ed., New York, American Institute of Physics, 1994, pp. 597 - 601.

[8] S. Humphries and J. Portillo, "Modelling relativistic electron beams with finite-element ray-tracing codes," in *Part. Accell. Conf.*, NY, USA, 1999.

[9] S. Humphies and J. Petillo, "Self-magnetic field calculations in ray-tracing codes," *Laser Part. Beams*, vol. 18, pp. 601 - 610, 2000.

[10] S. Humphries, Field solutions on computers, CRC Press, 1997.

[11] S. Humphries, "Technical information: TriComp Series," Field Precision, LLC, 2013. [Online]. Available: www.fieldp.com/technical.html.

[12] W. D. Stem, Y. J. Chen and J. L. Ellsworth, "Mitigation of nonlinear phase space in a space-charge-limited injector diode," Lawrence Livermore National Laboratory Report LLNL-PROC-788937 , 2019.

[13] T. L. Houck, Y. J. Chen and W. D. Stem, "Cathode misalignment simulations for ASD/Scorpius," Lawrence Laveremore National Laboratry Report LLNL-TR-813889, 2020.

[14] E. P. Lee and R. K. Cooper, "General envelope equation for cylindrically symmetric charged-particle beams," *Part. Accel.,* vol. 7, pp. 83-95, 1976.

[15] S. Humphries, Charged Particle Beams, New York: Wiley, 1990.

[16] M. Reiser, Theory and design of charged particle beams, New York. NY: Wiley, 1994.

[17] P. Allison, "Beam dynamics equations for XTR," Los Alamos National Laboratory report, LA-UR-01-6585, 2001.

[18] P. M. Lapostolle, "Possible emittance increase through filamentation due to space charge in continuous beams," *IEEE Trans. Nucl. Sci.,* vol. 18, pp. 1101 - 1104, 1971.

[19] K. L. Brown and et al., "TRANSPORT, a computerprogram for designing charged particle beam transport systems," CERN Report, CERN 80-04, 1980.

[20] P. Allison and et al., "Observation of self-steering effects on the ITS 6-MeV LINAC," in *Int. Part. Accel. Conf.*, Vancouver, BC, CA, 1997.

[21] C. Ekdahl and e. al., "First beam at DARHT-II," in *Part. Accel. Conf.*, 2003.

[22] C. Ekdahl, E. O. Abeyta, H. Bender, W. Broste, C. Carlson, L. Caudill, K. C. D. Chan, Y.-J. Chen, D. Dalmas, G. Durtschi, S. Eversole, S. Eylon, W. Fawley, D. Frayer, R. Gallegos, J. Harrison, E. Henestroza, M. Holzscheiter, T. Houck, T. Hughes, S. Humphries, D. Johnson, J. Johnson, K. Jones, E. Jacquez, B. T. McCuistian, A. Meidinger, N. Montoya, C. Mostrom, K. Moy, K. Nielsen, D. Oro, L. Rodriguez, P. Rodriguez, M. Sanchez, M. Schauer, D. Simmons, H. V. Smith, J. Studebaker, R. Sturgess, G. Sullivan, C. Swinney, R. Temple, C. Y. Tom and S. S. Yu, "Initial electron-beam results from the DARHT-II linear induction accelerator," *IEEE Trans. Plasma Sci.,* vol. 33, no. 2, pp. 892 - 900, 2005.

[23] C. A. Ekdahl, E. O. Abayta, P. Aragon and et al., "Long-pulse beam stability experiments on the DARHT-II linear induction accelerator," *IEEE Trans. Plasma Sci.,* vol. 34, pp. 460-466, 2006.

[24] C. Ekdahl, E. O. Abeyta, R. Archuleta, H. Bender, W. Broste, C. Carlson, G. Cook, D. Frayer, J. Harrison, T. Hughes, J. Johnson, E. Jacquez, B. T. McCuistian, N. Montoya, S. Nath, K. Nielsen, C. Rose, M. Schulze, H. V. Smith, C. Thoma and C. Y. Tom, "Suppressing beam motion in a long-pulse linear induction accelerator," *Phys. Rev. ST Accel. Beams,* vol. 14, no. 12, p. 120401, 2011.

[25] C. Ekdahl and et al., "Beam dynamics in a long-pulse linear induction accelerator," *J. Korean Phys. Soc.,* vol. 59, pp. 3448 - 3452, 2011.

[26] C. Ekdahl, "Tuning the DARHT long-pulse linear induction accelerator," *IEEE Trans. Plasma Sci.,* vol. 41, pp. 2774 - 2780, 2013.

[27] C. Ekdahl, "Modeling ion-focused transport of electron beams with simple beam-envelope simulations," Sandia National Laboratories Report SAND86-0544, 1986.

[28] T. P. Hughes, C. B. Mostrom, T. C. Genoni and C. Thoma, "LAMDA user's manual and reference," Voss Scientific Report, VSL-0707, 2007.

[29] Y.-J. Chen, "Corkscrew modes in linear induction accelerators," *Nucl. Instrum. Methods Phys. Res.,* vol. A292, no. no. 2, pp. 455 - 464, 1990.

[30] T. C. Genoni, T. P. Hughes and C. H. Thoma, "Improved envelope and centroid equations for high current beams," in *AIP Conf. Proc.*, 2002.

[31] T. C. Genoni, "Radial focusing of a relativistic electron beam in a bipotential electrostatic lens," *Phys. Rev. E,* vol. 50, no. 2, pp. 1496 - 1500, 1994.

[32] Y.-J. Chen, "Control of transverse motion caused by chromatic aberration and misalignments in linear accelerators," *Nucl. Instr. Meth. in Phys. Res. A,* vol. 398, pp. 139 - 146, 1997.

[33] V. K. Neil, L. S. Hall and R. K. Cooper, "Further theoretical studies of the beam breakup instability," *Part. Acc.,* vol. 9, pp. 213-222, 1979.

[34] G. J. Caporaso and Y. -J. Chen, "Electron Induction Linacs," in *Induction Accelerators*, K. Takayama and R. J. Briggs, Eds., New York, Springer, 2011, pp. 117 - 163.

[35] C. Ekdahl and R. McCrady, "Suppresion of beam breakup in linear induction accelerators by stagger tuning," *IEEE Trans. Plasma Sci.,* vol. 48, no. 10, pp. 3589 - 3599, 2020.

[36] W. K. H. Panofsky and M. Bander, "Asymptotic theory of beam-breakup in linear accelerators," *Rev. Sci. Instrum.,* vol. 39, pp. 206-212, 1968.

[37] G. Caporaso, "The control of beam dynamics in high energy induction linacs," in *Linear Accelerator Conf. (LINAC)*, 1986.

[38] C. Thoma and T. P. Hughes, "A beam-slice algorithm for transport of the DARHT-2 accelerator," in *Part. Acc. Conf.*, 2007.

[39] C. Ekdahl, "Tuning the DARHT Axis-II linear induction accelerator focusing," Los Alamos National Laboratory Report LA-UR-12-20808, 2012.

[40] C. Ekdahl and et al., "Emittance growth in linear induction accelerators," in *20th Int. Conf. High Power Part. Beams*, Washington, DC, USA, 2014.

[41] C. Ekdahl and M. Schulze, "Emittance growth in the DARHT Axis-II downstream transport," Los Alamos National Laboratory Technical Report, LA-UR-15-22706, 2015.

[42] C. Ekdahl and et al., "Emittance growth in the DARHT-II linear induction accelerator," *IEEE Trans. Plasma Sci.,*

vol. 45, no. 11, pp. 2962 - 2973, Nov 2017.

[43] C. Ekdahl, "Beam dynamics for ARIA," Los Alamos National Laboratory Report, LA-UR-14-27454, 2014.

[44] C. Ekdahl, "Electron-beam dynamics for an advanced flash-radiography accelerator," *IEEE Trans. Plasma Sci.,* vol. 43, no. 12, pp. 4123 - 4129, Dec. 2015.

[45] C. Ekdahl, "Beam Dynamics for Scorpius with the CDR end-to-end tune: I. Transport," Los Alamos National Laboratory Technical Report, LA-UR-18-, 2018.

[46] C. Ekdahl, "Beam breakup simulations for a solid state powered linear induction accelerator," Los Alamos National Laboratory Technical Report LA-UR-20-22662, Los Alamos, NM, USA, 2020.

[47] D. C. Carey, The optics of charged particle beams, New York: Harwood Academic, 1987, p. 99 et eq.

[48] D. Chernin, "Evolution of rms beam envelopes in transport systems with linear x-y coupling," *Part. Accel.,* vol. 24, pp. 29 - 44, 1988.

[49] T. C. Genoni and T. P. Hughes, "Ion-hose instability in a long-pulse linear induction accelerator," *Phys. Rev. - ST Accel. Beams,* vol. 6, no. 4, p. 030401, 2003.

[50] M. Weller, Interviewee, *Personal communication.* [Interview]. 26 January 2021.

[51] V. Kumar, D. Phadte and C. B. Patidar, "A simple formula for emittance growth due to spherical aberration in a solenoid lens," in *Proc. DAE-BRNS Indian Part. Accel. Conf.*, New Delhi, India, 2011.

[52] T. P. Wangler, K. R. Crandall, R. Ryne and T. S. Wang, "Particle-core model for transverse dynamics of beam halo," *Phys. Rev. Special Topics - Acc. Beams,* vol. 1, no. 8, p. 084201, 1998.

[53] H. V. Smith and et al., "X and Y offsets of the 18MeV DARHT-2 accelerator components inside the hall," Los Alamos National Laboratory report LA-UR-09-02040, 2009.

[54] H. V. Smith and et al., "X and Y tilts of the 18MeV DARHT-2 accelerator components," Los Alamos National Laboratory report LA-UR-09-03768, 2009.

[55] J. F. DeFord, "The AMOS (Azimuthal MOde Simulator) code," in *Proc. 13th Particle Accel. Conf.*, Chicago, IL, USA, 1989.

[56] L. Walling, P. Allison, M. Burns and et al., "Transverse impedance measurements of prototype cavities for a Dual-Axis Radiographic HydroTest (DARHT) facility," in *Proc. 14th Particle Accel. Conf.*, San Francisco, CA, USA, 1991.

[57] S. Kurennoy and R. McCrady, "Coupling Impedances of Ferrite-Loaded Cavities: Calculations and Measurements," in *11th Int. Part. Accel. Conf.*, Caen, FR, 2020.

[58] S. A. Heifets and S. A. Kheifets, "Coupling impedance in modern accelerators," *Rev. Mod. Phys.,* vol. 63, pp. 631 - 673, 1991.

[59] R. J. Briggs and W. Fawley, "Campaign to minimize the transverse impedance of the DARHT-2 induction linac cells," Lawence Berkeley National Laboratory Report LBNL-56796(Rev-1), 2002.

[60] Y. Tang, T. P. Hughes, C. A. Ekdahl and K. C. D. Chan, "BBU calculations for beam stability experiments on DARHT-2," in *European Part. Accel. Conf.*, Edinburgh, Scotland, 2006.